\documentclass[preprint,aps,amsmath,nofootinbib,amssymb,showkeys]{revtex4}
\usepackage{verbatim,graphics,graphicx,color,slashed,textcomp}
\usepackage{ulem} 
\usepackage[colorlinks=true,linkcolor=red,urlcolor=blue,citecolor=blue]{hyperref}

\graphicspath{{figures/}}

\newcommand{\GeV}{\ \mathrm{GeV}}
\newcommand{\TeV}{\ \mathrm{TeV}}
\newcommand{\nlsp}{\mathrm{NLSP}}
\newcommand{\lsp}{\mathrm{LSP}}	

\newcommand{\lrf}[2]{ \left(\frac{#1}{#2}\right)}
\newcommand{\lrfp}[3]{ \left(\frac{#1}{#2} \right)^{#3}}

\newcommand{\planck}{\texttt{Planck}}

\begin{document}

\hfill UT-17-19,  IPMU-17-0079

\title{Gravitino/Axino as Decaying Dark Matter and \\Cosmological Tensions}
\author{Koichi Hamaguchi$^{1,2}$, Kazunori Nakayama$^{1,2}$, Yong Tang$^{1}$}
\affiliation{${}^{1}$Department of Physics, Faculty of Science, University of Tokyo, Bunkyo-ku, Tokyo 113-0033, Japan
\\
${}^2$Kavli Institute for the Physics and Mathematics of the
 Universe (Kavli IPMU), University of Tokyo, Kashiwa 277--8583, Japan}
\date{\today}

\begin{abstract}
In supersymmetric axion models, if the gravitino or axino is the lightest SUSY particle (LSP),
the other is often the next-to-LSP (NLSP). We investigate the cosmology of such a scenario and point out that the lifetime of the NLSP naturally becomes comparable to the present age of the universe
in a viable parameter region. This is a well-motivated example of the so-called decaying dark matter model,
which is recently considered as an extension of the $\Lambda$CDM model to relax some cosmological tensions.

\end{abstract}
\maketitle

\section{Introduction}\label{sec:intro}

Supersymmetric (SUSY) version of the axion models have many good features: it solves the
gauge hierarchy problem, strong CP problem~\cite{Peccei:1977hh} and provides dark matter (DM) candidates
(see Ref.~\cite{Kawasaki:2013ae} for a review of SUSY axion model).
In particular, the existence of the axino, fermionic superpartner of the axion, makes the cosmological scenarios rich depending on its mass scale.
Although model-dependent, the axino mass tends to be the same order of the gravitino mass
in many realistic setups~\cite{Goto:1991gq, Chun:1992zk, Chun:1995hc,Kim:2012bb,Kawasaki:2013ae}.
Then it happens that if the gravitino or axino is the lightest SUSY particle (LSP), 
the other naturally becomes the next-to-LSP (NLSP).
In such a situation, the NLSP decays into the LSP plus axion, with a very long lifetime.
Although this decay channel is an ``invisible'' process in a sense that all the gravitino, axino and axion hardly interact with other particles, such an invisible decay can leave distinct features in cosmological observations.
We point out that the abundance of NLSP can be sizable and the lifetime of NLSP in this setup is naturally predicted to be close to the present age of the universe
for a reasonable scale of the Peccei-Quinn (PQ) symmetry breaking and the reheating temperature.
This is a well-motivated example of the decaying DM scenario, which is recently studied in the context of solving/relaxing cosmological tensions, explained below.\footnote{The gravitino (axino) decay into the axino (gravitino) and axion has been considered in some different motivations~\cite{Olive:1984bi,Chun:1993vz,Kim:1994ub,Asaka:2000ew,Ichikawa:2007jv,Hasenkamp:2011em,Graf:2013xpe}.
}

Since the discovery of our Universe's accelerated expansion, $\Lambda$CDM (cosmological constant/dark energy + cold dark matter) has been becoming the standard paradigm for cosmology. Various experiments have confirmed this simple picture by measuring key observables precisely in cosmic microwave background (CMB), large scale structure (LSS), expansion rate and so on. In spite of those successes, there are still some persistent tensions between different measurements. 
One of the most notable tensions is the Hubble constant $H_0$ measured by Hubble Space Telescope \cite{Riess:2016jrr} and \planck~\cite{Planck:2015xua}. The latest analysis~\cite{Riess:2016jrr} of HST data gives $H_0=73.24\pm 1.74$km s$^{-1}$Mpc$^{-1}$, which is about 3.4$\sigma$ higher than the value given by \planck~\cite{Planck:2015xua} within the $\Lambda$CDM model. Also, on the structure growth rate $\sigma _8$ (the amplitude of matter perturbation at scale around 8 Mpc), \planck ~data gives $\sigma _8=0.815\pm 0.009$, which is relatively larger than other low redshift measurements, for example, $\sigma_8(\Omega_m/0.27)^{0.46}=0.774\pm 0.040$ from weak lensing survey CFHTLenS~\cite{Heymans:2012gg}, and also about 2.3$\sigma$ deviation with the recent result from the KiDS-450 survey~\cite{Hildebrandt:2016iqg}. Those tensions deserve further investigations or improvements on the systematic uncertainties. On the other hand, they could also indicate extensions of the standard $\Lambda$CDM model and guide to new physics \cite{Pourtsidou:2016ico, DiValentino:2016hlg, Qing-Guo:2016ykt, Archidiacono:2016kkh,  Lesgourgues:2015wza, Murgia:2016ccp, Ko:2016uft, Ko:2016fcd, DiValentino:2016ucb, Barenboim:2016lxv, Chacko:2016kgg, Zhao:2017urm, Berezhiani:2015yta, Chudaykin:2016yfk, Enqvist:2015ara, Anchordoqui:2015lqa,Poulin:2016nat}.
In Refs.~\cite{Berezhiani:2015yta, Chudaykin:2016yfk}, a solution with decaying DM was proposed to resolve the above two tensions simultaneously by modifying the late-time cosmological evolution, where a sub-dominant component of DM ($\sim 5\%$) decays with lifetime $t_D\sim 10^{16}$s, shorter than the age of Universe. 
On the other hand, Ref.~\cite{Poulin:2016nat} claimed that the decaying DM does not drastically improve the situation, but still alleviates the tension in a right way especially for $t_D$ longer than the present age of the universe.
These observations may be regarded as a hint for decaying DM, or in the conservative viewpoint they just give constraint on the decaying DM scenario. In any case, the SUSY axion model is subject to these observations and it is worth studying its implications.
 
In Sec.~\ref{sec:ddm} we show that SUSY axion model naturally fits into the decaying DM scenario for reasonable range of parameters. In Sec.~\ref{sec:model} we construct an explicit model of SUSY axion which is suitable for the present purpose. Several cosmological constraints will also be discussed.

\section{Gravitino/Axino as decaying dark matter}\label{sec:ddm}

In SUSY theories to address the strong CP problem, axino, the fermionic superpartner of axion, is predicted~\cite{Rajagopal:1990yx}. 
Although the exact mass spectrum of axino $m_{\tilde{a}}$ depends on the details of model  which we shall come back later, some models could have $m_{\tilde{a}}$ of the order of gravitino mass $m_{3/2}$~\cite{Goto:1991gq, Chun:1992zk, Chun:1995hc, Kim:2012bb}. 
Thus, if the gravitino or axino is the LSP,  the other is often the NLSP, and the NLSP naturally becomes a long lived particle (decaying DM).

The interaction among the gravitino, axion and axino is
\begin{equation}
\mathcal{L}_{\textrm{int}}= -\frac{1}{2M_P}\partial_\nu a\bar{\psi}_\mu \gamma^\nu\gamma^\mu i\gamma_5 \tilde{a},
\end{equation}
where $M_P=2.4\times 10^{18}\GeV$, $\tilde{a}$ is axino, $a$ is the axion, and $\psi_\mu$ is the gravitino. 
We consider two scenarios; 
(i) the gravitino NLSP $+$ the axino LSP, and 
(ii) the axino NLSP $+$ the gravitino LSP.
\begin{itemize}
\item[(i)] In the models with the gravitino NLSP and the axino LSP, we obtain\footnote{
	Actually there may be a relative phase between the gravitino mass and axino mass in the model basis.
	This phase affects the decay rate through the term which is linear in $r_{\tilde a}$. 
}
\begin{equation}\label{fig:gdecayw}
\Gamma(\psi_\mu \rightarrow \tilde{a} + a)=\frac{m^3_{3/2}}{192\pi M^2_P}
(1-r_{\tilde{a}})^2
(1-r_{\tilde{a}}^2)^3, \qquad
r_{\tilde{a}} \equiv\frac{m_{\tilde{a}}}{m_{3/2}},
\end{equation}
where $m_{3/2}$ and $m_{\tilde a}$ denote the gravitino and axino mass respectively and
we have neglected the axion mass. When $m_{\tilde{a}}\ll m_{3/2}$ the decay width or lifetime of gravitino goes as
\begin{equation}\label{fig:adecayw}
\Gamma^{-1}\simeq \frac{192\pi M_P^2}{m_{3/2}^3}
\simeq 2.35\times 10^{15}\text{sec}\cdot \left(\frac{1\GeV}{m_{3/2}}\right)^{3}.
\end{equation}

\item[(ii)]
In the models with the axino NLSP and the gravitino LSP, we obtain
\begin{align}
\Gamma(\tilde{a} \rightarrow  \psi_\mu + a)=\frac{m_{\tilde{a}}^5}{96\pi m^2_{3/2} M^2_P}
(1-r_{\tilde{a}}^{-1})^2
(1-r_{\tilde{a}}^{-2})^3.
\end{align}

\end{itemize}
The above decay widths agree with~\cite{Chun:1993vz,Kim:1994ub,Asaka:2000ew} in the limit of massless final states.

Now let us discuss the abundances of the gravitino and the axino.
The production rate of gravitino depends on the temperature of thermal bath. The relic abundance of gravitino is given by Refs.~\cite{Bolz:2000fu, Pradler:2006qh, Rychkov:2007uq}
\begin{equation}\label{eq:gravitinoDM}
\Omega_{3/2}h^2\simeq 
0.02 \lrf{T_R}{10^5\GeV} \lrf{1\GeV}{m_{3/2}} \lrfp{M_3(T_R)}{3\TeV}{2} \lrf{\gamma(T_R)/(T_R^6/M_P^2)}{0.4},
\end{equation}
where $T_R$ is the reheating temperature after the inflation, $M_3(T_R)$ is the running gluino mass, and the last factor parametrizes the effective gravitino production rate whose value changes as $\gamma(T_R)/(T_R^6/M_P^2)\simeq 0.4$--0.35 for $T_R\simeq 10^4$--$10^6\GeV$~\cite{Rychkov:2007uq}. Thermal gravitino production occurs mostly by the scattering processes of particles with strong interaction. Since the production is UV-dominant, higher reheating temperature gives larger relic abundance. Here we neglect the possible non-thermal contribution from, for example, other SUSY particle's decay, which are highly model dependent.  In the numerical calculation, we use an approximation $M_3(T_R) = m_{\tilde{g}}\times (g_3(T_R)/g_3(m_{\tilde{g}}))^2$, where $m_{\tilde{g}}$ is the gluino mass and $g_3(\mu)$ is the running SU(3) coupling.

The axino production may be dominated by the scatterings of gluons and gluinos. In such a case,
the axino relic density depends on the axino mass $m_{\tilde{a}}$, the reheating temperature $T_R$ and the PQ symmetry breaking scale $f_a$~\cite{Brandenburg:2004du,Strumia:2010aa}, 
\begin{equation}\label{eq:axinoDM}
\Omega_{\tilde{a}}h^2\simeq 0.30\times g_3(T_R)^4 \lrf{F(g_3(T_R))}{23} \lrf{m_{\tilde{a}}}{1\GeV} \lrf{T_R}{10^4\GeV} \lrfp{10^{12}\GeV}{f_a}{2},
\end{equation}
where the function $F$ changes as $F\simeq  24$--21.5 for $T_R\simeq 10^4$--$10^6\GeV$~\cite{Strumia:2010aa}.
Here we assumed the KSVZ axion model~\cite{Kim:1979if,Shifman:1979if}.
In the DFSZ model~\cite{Dine:1981rt, Zhitnitsky:1980tq}, the axino production would be dominated by the axino-Higgs-higgsino
and/or axino-quark-squark interactions and becomes independent of the reheating temperature~\cite{Chun:2011zd,Bae:2011jb}.

To be complete, we also show the axion contribution to DM from the misalignment production~\cite{Turner:1985si}, 
\begin{equation}\label{eq:axionDM}
\Omega_a h^2\simeq 0.18\theta_i^2 \lrfp{f_a}{10^{12}\GeV}{1.19},
\end{equation}
where $\theta_i f_a$ is the initial value of the misaligned axion field. 
Since we are interested in relatively low reheating temperature $(T_R \ll f_a)$, we naturally expect that the PQ symmetry is not restored during/after inflation.
Thus we do not need to take care of the axion production from topological defects.

For convenience, we define the following useful quantity $F$:
\begin{equation}
\Omega_{\nlsp}h^2 \frac{m_{\nlsp}-m_{\lsp}}{m_{\nlsp}}\simeq F \Omega_{\rm cdm}h^2,   \label{ddmF}
\end{equation}
where $\Omega_{\rm cdm}h^2 \simeq 0.12$ and $F$ parametrizes the fraction of cold dark matter that is transferred into radiation. This definition also contains the case where $m_{\nlsp}$ is close to $m_{\lsp}$ and the LSP remains non-relativistic. 

\begin{figure}[thb]
	\includegraphics[width=0.6\textwidth,height=0.6\textwidth]{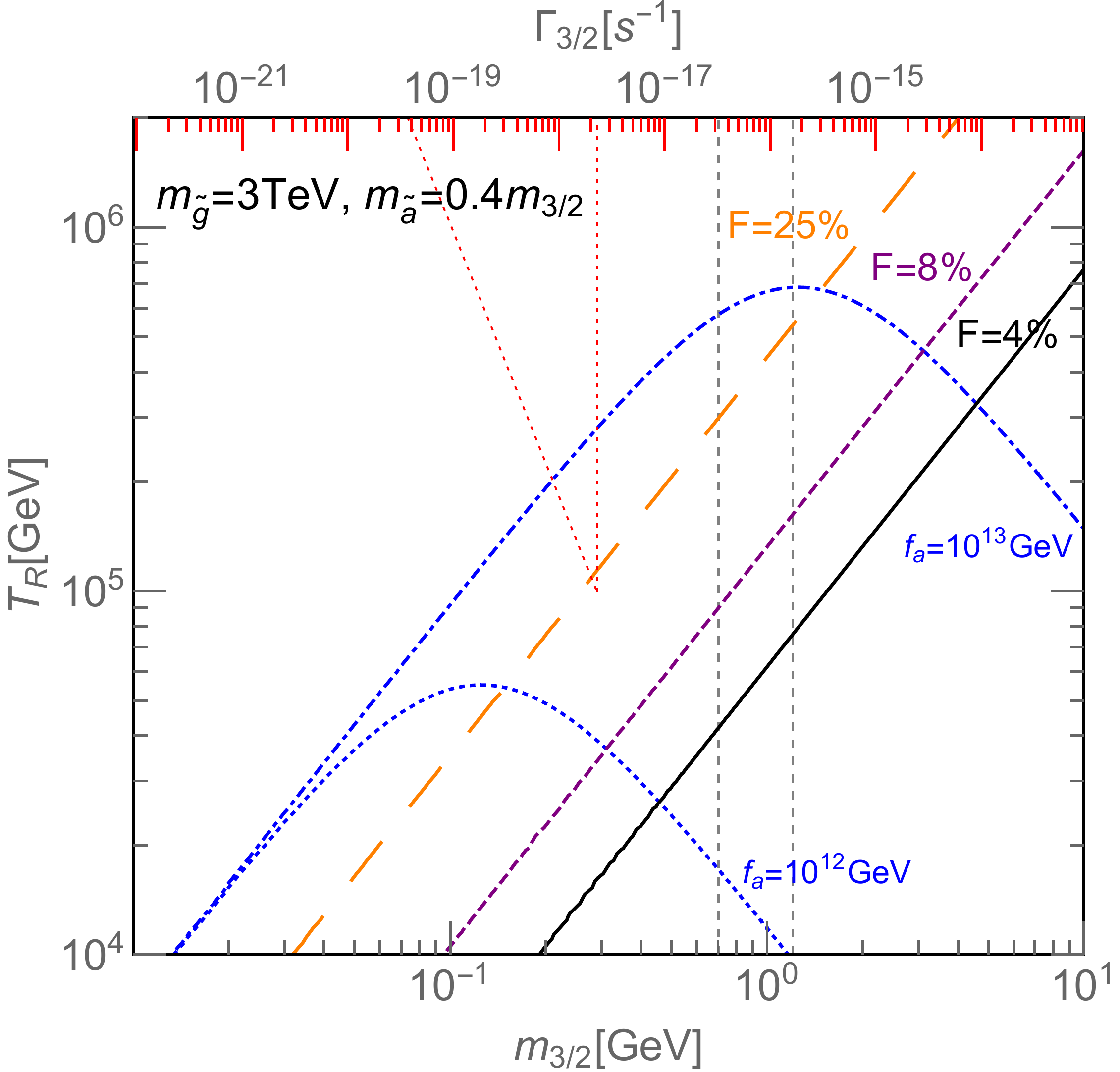}
	\caption{The relevant contours for the gravitino NLSP scenario in the plane of the gravitino mass $m_{3/2}$ vs the reheating temperature $T_R$, for $m_{\tilde{g}}=3\TeV$ and $m_{\tilde{a}}=0.4m_{3/2}$. Contours of $\Omega_{3/2}h^2+\Omega_{\tilde{a}}h^2\simeq 0.12$ are shown in blue dotted and dot-dashed curves for PQ scales $f_a=10^{12}$ and $10^{13}\GeV$, respectively. Contours of the decaying DM fraction, $F=4\%$, $8\%$ and $25\%$, are shown in solid black, dashed purple and long-dashed orange lines, respectively. Region between the two vertical gray dashed lines may be interesting from the viewpoint of cosmological tensions~\cite{Berezhiani:2015yta, Chudaykin:2016yfk}. Region inside the red dotted triangle in the upper part indicates the exclusion from Ref.~\cite{Poulin:2016nat}. \label{fig:mgtr}}
\end{figure}

In Fig.~\ref{fig:mgtr}, we show the results for the gravitino NLSP scenario in the plane of $m_{3/2}-T_R$, 
with gluino mass $m_{\tilde{g}}=3\TeV$ and $m_{\tilde{a}}=0.4m_{3/2}$. 
The blue dotted and dot-dashed curves give $\Omega_{3/2}h^2+\Omega_{\tilde{a}} h^2 \simeq 0.12$ with $f_a=10^{12}$ and $10^{13}\GeV$, respectively. For fixed $f_a$, regions below the blue curves are allowed since the axion contribution to DM can compensate with proper initial $\theta_i$ in Eq.~(\ref{eq:axionDM}). 
We also show the contours of the decaying DM fraction $F=4\%$, $8\%$ and $25\%$ in solid black, dashed purple and long-dashed orange lines, respectively, which illustrates how the gravitino contribution shifts in the $m_{3/2}-T_R$ plane. 
Notably, within the allowed parameter ranges the gravitino lifetime is rather close to the present age of the universe $t_0\simeq 4\times 10^{17}\text{s}$, 
and also its energy fraction to the total DM is sizable. Therefore, the NLSP gravitino naturally becomes a good candidate of the decaying DM.
The parameter region between the two vertical gray dashed lines may be interesting from the viewpoint of cosmological tensions~\cite{Berezhiani:2015yta, Chudaykin:2016yfk}, $\Gamma \simeq 3.2\times 10^{-17}\sim 1.6\times 10^{-16}$s$^{-1}$ and $F \sim $ a few percent.
In addition, we also show the constraint on the decaying DM from Refs.~\cite{Poulin:2016nat} for $\Gamma < H_0$ with red dotted lines in the upper part of the figure. The region inside this dotted triangle is excluded by combination of various cosmological data. 
Conversely, the region near this red dotted line may be favored since the cosmological tensions are relaxed.
For the case of $\Gamma > H_0$, Ref.~\cite{Poulin:2016nat} obtained an upper bound on $F\lesssim 4\%$, which coincides with our solid black line and, if robust, would exclude the region above.

\begin{figure}[t]
	\includegraphics[width=0.6\textwidth,height=0.6\textwidth]{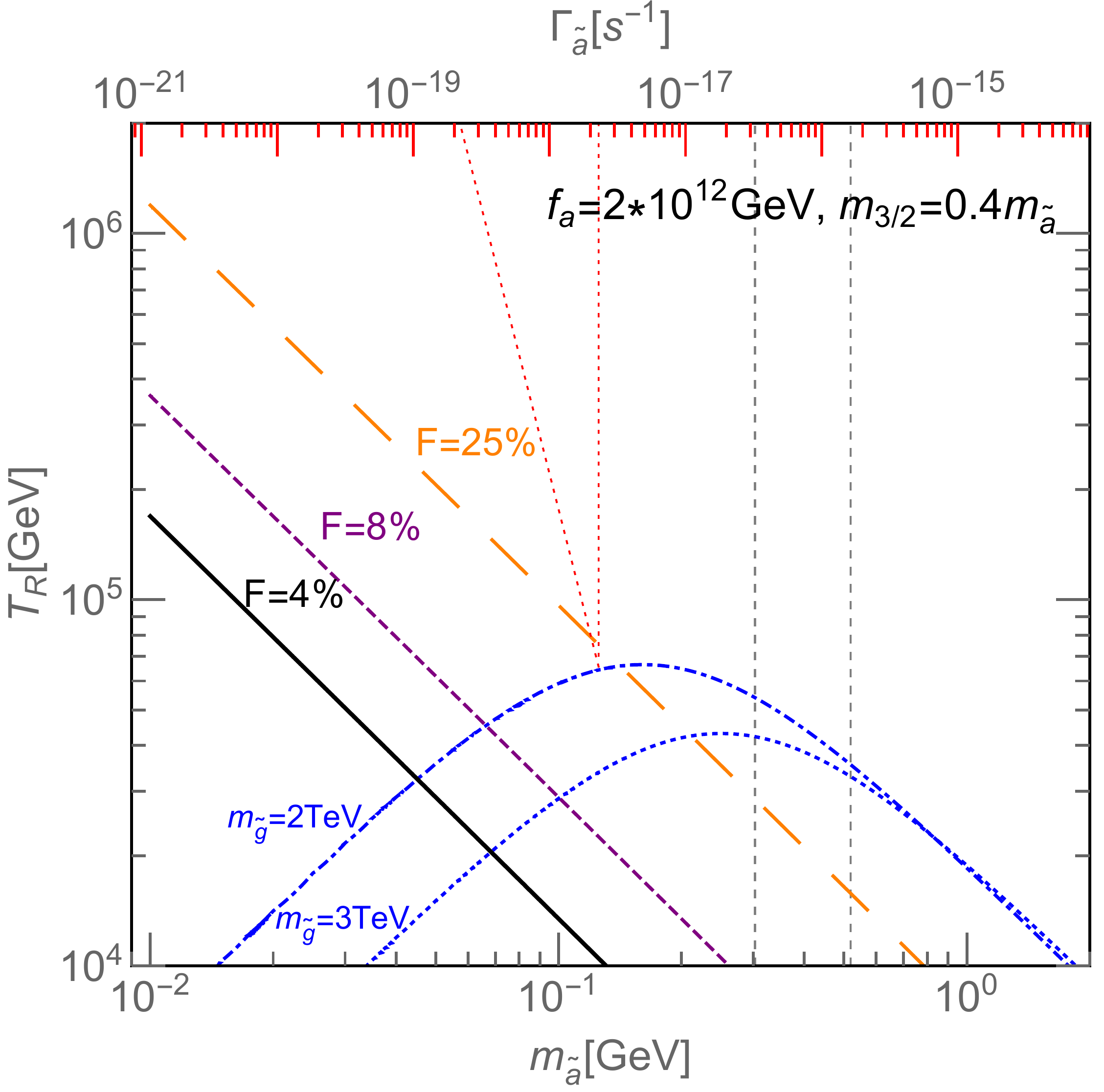}
	\caption{The relevant contours for the axino NLSP scenario in the plane of the axino mass $m_{\tilde{a}}$ vs the reheating temperature $T_R$, for $f_a=2\times10^{12}\GeV$ and $m_{3/2}=0.4m_{\tilde{a}}$. Contours of $\Omega_{3/2}h^2+\Omega_{\tilde{a}}h^2\simeq 0.12$ are shown in blue dot-dashed and dotted curves for the gluino mass $m_{\tilde{g}}=2$ and $3\TeV$, respectively.
Other lines are the same as Fig.~\ref{fig:mgtr}. \label{fig:matr}}
\end{figure}

Fig.~\ref{fig:matr} shows the case of the axino NLSP scenario, with $f_a=2\times 10^{12}\GeV$ and $m_{3/2}=0.4m_{\tilde{a}}$.
In this case, a relatively lower reheating temperature $T_R$ is necessary to give a viable DM relic density, which can be understood by comparing the mass scales in Eqs.~\eqref{eq:gravitinoDM} and \eqref{eq:axinoDM}. For fixed $m_{\tilde{g}}$, regions below the blue curves are allowed. As can  be seen in the figure, there is a parameter space between the two vertical gray dashed lines that might be interesting for cosmological tensions. Again, the constraint from Ref.~\cite{Poulin:2016nat} is shown by the red dotted triangle. 

We emphasize that the allowed parameter regions are not so large in both cases and hence it is the \textit{prediction} of the 
gravitino/axino LSP scenario that the NLSP has a lifetime comparable to the present age of the universe, as often considered in the context of decaying DM model
to relax the cosmological tensions.

\section{A model of SUSY axion}\label{sec:model}

Now we provide one of the concrete setups of a SUSY axion model and its cosmological implications. 
Let us consider a SUSY axion model with superpotential
\begin{align}
	W = \lambda X(\phi\bar\phi - f^2) + W_\phi +  W_0,  \label{WPQ}
\end{align}
where $\phi$ and $\bar\phi$ are PQ scalars with global U(1) charges of $+1$ and $-1$ while $X$ is neutral under the U(1),
$W_0 = m_{3/2}M_P^2$ is a constant term and $W_\phi$
involves the coupling of $\phi$ with some other sector to make the global U(1) anomalous under the QCD.
In the KSVZ model we have $W_\phi = y\phi Q\overline Q$ with $Q$ and $\overline Q$ denoting the additional vector-like quarks and in the DFSZ model we have $W_\phi = \kappa\phi^2 H_u H_d/M_P$ with $H_u$ and $H_d$ denoting the up- and down-type Higgs doublets, which may also solve the so-called $\mu$ problem~\cite{Kim:1983dt}, although we focus on the KSVZ model in this letter.
We also assume gauge-mediated SUSY breaking model~\cite{Giudice:1998bp} to make the gravitino mass much smaller than the soft SUSY breaking scale.

Without loss of generality, all the parameters in Eq.~(\ref{WPQ}) are taken to be real and positive by field redefinition. 
Including the SUSY breaking effects, the scalar potential is given by
\begin{align}
	V = m_\phi^2|\phi|^2 + m_{\bar\phi}^2|\bar\phi|^2 + \lambda^2\left(\left|\phi\bar\phi - f^2\right|^2+|X|^2(|\phi|^2+|\bar\phi|^2)\right) + 2\lambda m_{3/2}f^2(X+X^\dagger). 
\end{align}
The soft masses are naively $m_\phi\sim m_{\bar\phi}\sim \mathcal O(m_{3/2})$ from gravity mediation effects.
In the KSVZ model, $\phi$ couples to $Q$ and $\overline Q$ and hence it also receives soft masses from gauge mediation effect.
This gives negative contribution to the soft masses as ${m_{\phi}^{2}}^{\rm (GMSB)} \sim - y^2 m_{\rm soft}^2/\pi^2$ ~\cite{Asaka:1998ns,ArkaniHamed:1998kj,Choi:2011rs,Nakayama:2012zc}.
If this is dominant, we need some extra potential to stabilize the PQ field at appropriate field value.
In the present analysis, we neglect it by assuming that $y$ is small enough. 
The potential minimum is
\begin{align}
	&\left<X\right> = -\frac{2m_{3/2}f^2/\lambda}{\left<|\phi|\right>^2+\left<|\bar\phi|\right>^2},\\
	&\left<|\phi|\right> = f \left( \frac{m_{\bar\phi}^2 + \lambda^2\left< X\right>^2}{m_{\phi}^2 + \lambda^2\left< X\right>^2}\right)^{1/4} 
	+ \mathcal O\left(\frac{m_\phi^2}{f}\right),\\
	&\left<|\bar\phi|\right> = f \left( \frac{m_{\phi}^2 + \lambda^2\left< X\right>^2}{m_{\bar\phi}^2 + \lambda^2\left< X\right>^2}\right)^{1/4}
	+ \mathcal O\left(\frac{m_{\bar\phi}^2}{f}\right).
\end{align}
The PQ scale $f_a$ which appeared in the previous section is given by $f_a = \sqrt{2 \left(\left<|\phi|^2\right> + \left<|\bar\phi|^2\right> \right)}/N_{\rm DW}$
where $N_{\rm DW}$ denotes the domain wall number which is equal to unity in the simplest KSVZ axion model.
Note that the phase of the product $\phi\bar\phi$, or $\arg(\phi)+\arg(\bar\phi)$, is determined by minimizing the potential to be zero,
but $\arg(\phi)-\arg(\bar\phi)$ remains undetermined, corresponding to the axion.
This freedom allows us to redefine the phases so that $\arg(\phi)=\arg(\bar\phi)=0$.
The mass matrix of the fermionic components are given by
\begin{align}
	\mathcal L = -\frac{1}{2}\left( \widetilde X,~\widetilde\phi,~\widetilde{\bar\phi} \right) \mathcal M 
	\begin{pmatrix}
		\widetilde X \\ \widetilde\phi \\ \widetilde{\bar\phi}
	\end{pmatrix}+ {\rm h.c.},~~~~~~
	\mathcal M = \lambda\begin{pmatrix}
		0 &\left<\bar\phi\right> &\left<\phi\right> \\ \left<\bar\phi\right> & 0 & \left<X\right> \\ \left<\phi\right> & \left<X\right> & 0
	\end{pmatrix}.
\end{align}
Since $\left<\phi\right> ,\left<\bar\phi\right> \gg \left<|X|\right>$, the two heavy mass eigenvalues are both found to be
$m_{\rm heavy} \simeq \lambda\sqrt{ \left<\phi\right>^2+\left<\bar\phi\right>^2 }$.
Noting that $\det({\mathcal M}) = 2\lambda^3\phi\bar\phi X\simeq 2\lambda^3 f^2 X$, the light mass eigenvalue is given by
\begin{align}
	m_{\tilde a} = \left|\frac{2\lambda f^2 \left<X\right>}{ \left<\phi\right>^2+\left<\bar\phi\right>^2}\right| = \frac{4f^4m_{3/2}}{ \left(\left<\phi\right>^2+\left<\bar\phi\right>^2\right)^2},
\end{align}
This light state is identified with axino. Neglecting corrections of $\mathcal O(m_{3/2}/f)$, the axino mass eigenstate is
\begin{align}
	\widetilde a = \frac{1}{\sqrt{ \left<\phi\right>^2+\left<\bar\phi\right>^2}}\left(- \left<\phi\right>\widetilde\phi + \left<\bar\phi\right>
	\widetilde{\bar\phi} \right).
\end{align}
In general, $\left<\phi\right> \sim \left<\bar\phi\right> \sim f$ and hence $m_{\tilde a} \sim m_{3/2}$.
Thus the gravitino mass is comparable to the axino mass in this model.
It is possible to make a slight hierarchy between the gravitino and axino mass 
if either $m_{\phi}^2 > m_{\bar\phi}^2$ or $m_{\phi}^2 < m_{\bar\phi}^2$
so that $ \left<\bar\phi\right> > f > \left<\phi\right>$ or $ \left<\bar\phi\right> < f < \left<\phi\right>$.\footnote{
In the KSVZ model, the axino may radiatively obtain a mass from the gauge mediation effect as
$m_{\tilde a}^{\rm (GMSB)} \sim y^2 A_{\phi}/\pi^2$ with $A_\phi$ being the soft SUSY breaking $A$-term related to the superpotential $W_\phi$.
This contribution is small enough once we assume that the gauge-mediation contribution to the saxion mass is small enough.
}

Let us briefly consider the saxion cosmology. 
Because of the holomorphic property of the superpotential, the global U(1) indicates the existence of the flat direction in the scalar potential,
which is only lifted up by the SUSY breaking effect and it is identified with the saxion.
In the model presented above, there is a moduli space along $\phi\bar\phi=f^2$ and the mass of this direction is of the order of $m_{3/2}$.
The saxion begins a coherent oscillation when the Hubble expansion rate becomes comparable to the saxion mass, $H\sim m_{3/2}$, with amplitude of $\sim f$. Thus, its the abundance is\footnote{
For the KSVZ model, thermal effects may be marginally important for interesting parameter regions~\cite{Kawasaki:2010gv,Kawasaki:2011ym,Moroi:2012vu}.}
\begin{align}
	\frac{\rho_s}{s} \sim \frac{T_R}{8}\left( \frac{\theta f}{M_P} \right)^2 \sim 
	2\times 10^{-9}\,{\rm GeV}\left( \frac{T_R}{10^5\,{\rm GeV}} \right)\left( \frac{f}{10^{12}\,{\rm GeV}} \right)^2 \theta^2,
	\label{rhos}
\end{align}
where $\theta \sim \mathcal O(0.1)$ represents the uncertainly of the initial amplitude.
Saxions are also produced thermally by the scatterings of particles in thermal bath, 
but it is subdominant or at most comparable to the coherent oscillation for the parameters of our interest.
The lifetime of the saxion (assuming that it mainly decays into the axion pair) is given by~\cite{Chun:1995hc,Graf:2012hb}
\begin{align}
	\tau_s \simeq \left( \frac{x^2 m_s^3}{32\pi f_a^2 N_{\rm DW}^2} \right)^{-1} 
	\sim 2\times 10^2\,{\rm sec}\, x^{-2}\left( \frac{1\,{\rm GeV}}{m_s} \right)^3\left( \frac{f_a N_{\rm DW}}{10^{12}\,{\rm GeV}} \right)^2,
	\label{taus}
\end{align}
where $x \equiv \left(\left<|\phi|^2\right> - \left<|\bar\phi|^2\right> \right) / \left(\left<|\phi|^2\right> + \left<|\bar\phi|^2\right> \right)$
is in general an $\mathcal O(0.1)$ numerical coefficient.
Therefore it may decay after the Big-Bang Nucleosynthesis (BBN) begins and hence we must take care of the BBN constraint. 
Although the saxion dominantly decays into the axion pair, a small fraction goes into the visible sector.
For the saxion mass of $\sim 1$\,GeV, it may decay into two photons and two mesons.
First, the branching fraction of two-photon decay is about $B_r\sim \alpha_{\rm EM}^2N_{\rm DW}^2 /(8\pi^2x^2)\sim 10^{-6}$ with $\alpha_{\rm EM}$ being electro-magnetic fine structure constant. It is harmless for this mass range since the constraint reads $B_r(\rho_s/s) \lesssim 10^{-9}$\,GeV for $\tau_s \sim 10^4$\,sec, for example
~\cite{Kawasaki:2004qu}. (See also Ref.~\cite{Kawasaki:2007mk} for more detailed constraints in a similar setup.)
Second, the branching fraction of two-meson decay is expected to be $B_h\sim \alpha_s^2N_{\rm DW}^2 /(\pi x)^2\sim 10^{-3}$ with $\alpha_s$ being the QCD fine structure constant.
According to Ref.~\cite{Pospelov:2010cw}, the constraint roughly reads $B_h(\rho_s/s)/m_s \lesssim 10^{-10}-10^{-9}$
for $\tau_s \simeq 10^2-10^5$\,sec.
This is also not a strong constraint in the parameter ranges of our interest.
Let us also comment on the constraint from the axion dark radiation abundance.
It requires that the extra effective number of neutrino species should be smaller than $\mathcal O(0.1)$~\cite{Planck:2015xua},
which roughly means that the saxion must decay before it dominates the universe~\cite{Kawasaki:2007mk,Graf:2013xpe}. 
From Eq.(\ref{rhos}) and Eq.(\ref{taus}), it is well satisfied in the parameter ranges of our interest.

Finally, the lightest SUSY particle except for the gravitino/axino (stau, for example) can decay into the gravitino/axino that may be able to affect BBN. For the stau mass $m_{\tilde\tau}$ larger than $200\GeV$ and $m_{3/2}\sim 1\GeV$, the BBN bound is safely evaded~\cite{Kawasaki:2008qe}.

\section{Conclusions}\label{sec:conc}

We have shown that SUSY axion models naturally predict the gravitino/axino LSP-NLSP system, 
and the lifetime of the NLSP becomes of the order of the present age of the universe for natural parameter ranges of
the PQ breaking scale and the reheating temperature.
This is a good example of the so-called decaying DM scenario, which is recently introduced 
in the context of tensions between cosmological parameters deduced from the CMB observation and low-redshift astronomical observations.
Taking the tension seriously, the decaying DM scenario is one of the options as an extension of the $\Lambda$CDM model 
and hence it is worth studying whether such a model is natural or not from the particle physics point of view.

If the cosmological tension is really explained by the gravitino/axino decaying DM scenario,
there are several observable predictions.
First, the PQ scale is close to $10^{12}$\,GeV and the axion coherent oscillation is expected to constitute sizable fraction of DM.
It is within the reach of the cavity search experiment~\cite{Sikivie:1983ip,Bradley:2003kg,Asztalos:2009yp}.
The future of proposed experiment to measure the effect of long-range force mediated by the axion may also be sensitive to this axion mass scale~\cite{Arvanitaki:2014dfa}.
Second, the lightest SUSY particle except for the gravitino/axino is long-lived at the collider scale.
If it is the stau, charged tracks may be observed at the LHC. The current bound~\cite{CMS:2016ybj} is $m_{\tilde{\tau}}\gtrsim 360\GeV$ and the future data at $14\TeV$ LHC might probe up to $m_{\tilde{\tau}}\simeq 1.20\TeV$~\cite{Feng:2015wqa}. Thus the discovery of both the axion and long-lived SUSY particle signatures would test or support our scenario.

\section*{Acknowledgments}

This work was supported by the Grant-in-Aid for Scientific Research on Scientific
Research A (No.26247038 [KH], No.26247042 [KN], No.16H02189 [KH]), Young Scientists B
(No.26800121 [KN], No.26800123 [KH]) and Innovative Areas (No.26104001
[KH], No.26104009 [KH and KN], No.15H05888 [KN], No.16H06490 [YT]), and
by World Premier International Research Center Initiative (WPI
Initiative), MEXT, Japan.


\providecommand{\href}[2]{#2}\begingroup\raggedright\endgroup

\end{document}